\documentstyle[prl,aps,epsf,epsfig]{revtex}

\def\la{\lambda}
\def\ra{\rightarrow}
\def\k{m_{eff}}

\def\p{\pi}
\def\th{\theta}

\def\bar{\overline}
\def\l{\lambda}

\newcommand{\AmS}{{\protect\the\textfont2
  A\kern-.1667em\lower.5ex\hbox{M}\kern-.125emS}}

\def\bentarrow{\:\raisebox{1.1ex}{\rlap{$\vert$}}\!\rightarrow}

\def\dk#1#2#3{
        \begin{equation}
        \begin{array}{r c l}
        #1 & \rightarrow & #2 \\
         & & \bentarrow #3
        \end{array}
        \end{equation}
                }
\def\dkdk#1#2#3#4{
        \begin{equation}
        \begin{array}{r c l}
        #1 & \rightarrow & #2 \\
         & & \bentarrow #3 \\
         & & \phantom{\;\bentarrow}\bentarrow #4
        \end{array}
        \end{equation}
                }
\def\dkp#1#2#3#4{
        \begin{equation}
        \begin{array}{r c l}
        #1 & \rightarrow & #2#3 \\
         & & \phantom{\; #2}\bentarrow #4
        \end{array}
        \end{equation}
                }

\def\beq{ \begin{equation}}
\def\eeq{\end{equation} }
\def\bea{\begin{eqnarray}}
\def\eea{\end{eqnarray}}

\begin{document}
\draft
\twocolumn[\hsize\textwidth\columnwidth\hsize\csname
@twocolumnfalse\endcsname

\title{
\begin{flushright}
{\large \tt CERN-TH/99-40 \\ 
hep-ph/9903203}
\end{flushright}
\rule{16.3cm}{0.2mm}
{\Large \bf 
Aspects of Neutrino Masses and Lepton-Number Violation  \\
in the light of the Super-Kamiokande data
} 
\rule{16.3cm}{0.2mm} 
}

\author{{\large \bf Smaragda Lola} \\
\vspace*{0.2 cm}
{\large  Theoretical Physics Division, CERN} \\
{\large CH-1211 Geneva 23, Switzerland} \\
\vspace*{0.45 cm}
{\normalsize \tt Contribution to
the 1998 Corfu Summer Institute  on \\
Elementary Particle Physics (JHEP proceedings)
}}

\maketitle
.

\begin{abstract}
 {\normalsize {\large \bf Abstract:}  
We discuss aspects of neutrino masses and lepton-number violation, in the
light of the observations of Super-Kamiokande. As 
a first step, we use the data from various experiments, 
in order to obtain a phenomenological understanding of 
neutrino mass textures. We then investigate how the required
patterns of neutrino masses and mixings are related to the 
flavour structure of the underlying theory. In 
supersymmetric extensions of the Standard Model, renormalisation 
group effects can have important implications: for small 
$\tan\beta$, $b$--$\tau$ unification indicates the presence of 
significant $\mu$--$\tau$ flavour mixing.  The evolution of 
the neutrino mixing 
may be described by 
simple semi-analytic expressions, which confirm that,
for large $\tan\beta$, 
very small mixing at the GUT scale may be amplified to 
maximal mixing at low energies, and vice versa. 
Passing to 
specific models, we first discuss the predictions for 
neutrino masses in different GUT models (including 
superstring-embedded solutions). Imposing the requirement for
successful leptogenesis may give additional constraints on
the generic structure of the neutrino mass textures.
Finally, we discuss direct ways 
to look for lepton-number violation in ultra-high energy neutrino 
interactions. \\
\vspace*{0.4 cm} }
\end{abstract}
]

\begin{narrowtext}

\section{ Neutrino data and Implications}

\noindent
Recent reports by Super-Kamiokande \cite{SKam} 
and other experiments \cite{KamMac}, support
previous measurements of a 
$\nu _{\mu }/\nu _{e }$ ratio
in the atmosphere that is significantly 
smaller than the theoretical expectations. 
The data favours
$\nu _{\mu }$--$\nu _{\tau }$ oscillations, with 
\begin{eqnarray}
\delta m_{\nu _{\mu }\nu _{\tau }}^{2} &\approx &(10^{-2}\;{\rm to}%
\;10^{-3})\;{\rm eV^{2}} \\
\sin^{2}2\theta _{\mu \tau } &\geq &0.8  \label{atmos}
\end{eqnarray}
(Oscillations involving a sterile neutrino
are also plausible, while dominant $\nu _{\mu }\rightarrow \nu _{e}$
oscillations are disfavoured by Super-Kamiokande~\cite{SKam} and CHOOZ ~\cite
{chooz}.).

On the other hand, the solar neutrino puzzle can be resolved through
either
vacuum or matter-enhanced (MSW) oscillations. The first require a mass
splitting of the neutrinos
that are involved in the oscillations 
in the range
$\delta m_{\nu _{e}\nu _{\alpha }}^{2} 
\approx (0.5-1.1)\times 10^{-10}~{\rm eV}^2$,
where $\alpha $ is $\mu $ or $\tau$. 
MSW oscillations on the other hand \cite{MSW}, allow for both small
and large mixing, while now
$ \delta m_{\nu _{e}\nu _{\alpha }}^{2} \approx (0.3-20) \times 10^{-5}
~{\rm eV}^2$.
Moreover, the  LSND collaboration 
has reported evidence for the appearance of $\bar{\nu}%
_{\mu }$--$\bar{\nu}_{e}$ and ${\nu }_{\mu }$--${\nu }_{e}$
oscillations \cite{LSND2}, which however are not supported by KARMEN 2 
\cite{Karmen2}.
Finally, if neutrinos were to provide a hot dark matter component, then the
heavier neutrino(s) should have mass in the range $\sim (1-6)$ eV.

The implications of these observations are very interesting, since
they  point towards a non-zero neutrino mass and lepton-number 
violation, that is
{\em the existence of physics  beyond
the standard model}. The simplest class of solutions
that one may envisage  consist of an
extension of the Standard Model to include three new
right-handed neutrino states, with a 
mass structure directly related to that of the other fermions.
Three neutrino masses
allow only two independent mass differences and thus
the direct indications for neutrino oscillations discussed
above cannot be simultaneously explained unless a 
sterile light-neutrino state is introduced.
Since the LSND results have not been confirmed, in our
analysis we chose not to introduce sterile states
(which inevitably break any simple connection of the neutrino
masses with the known charge lepton and quark hierarchies).
Instead, we focus on the
Super-Kamiokande and the solar neutrino data, and leave
open the possibility of neutrinos as hot dark matter.

In this framework, both the solar and atmospheric
deficits  require small mass differences, and thus
can be explained by two possible neutrino hierarchies:

(a) Textures with almost 
degenerate neutrino eigenstates, with mass ${\cal O} 
({\rm eV})$.  In this case neutrinos may
also provide a component of hot dark matter.

(b) Textures with 
large hierarchies of neutrino masses:
$m_{3} \gg m_{2}, m_{1}$,
with the
possibility of a second hierarchy
$m_{2} \gg m_{1}$. 
Then, the atmospheric neutrino data
require
$m_{3} \approx (10^{-1} \; {\rm to} \;
10^{-1.5})$ eV
and $m_{2} \approx (10^{-2} \; {\rm to} \;
10^{-3})$ eV.

What is the information we can obtain from
these data on the underlying neutrino structure?
To answer this, one starts by describing the
most general neutrino mass matrix for three flavours 
of isodoublet and isosinglet neutrinos.
This matrix, in the current eigenstate basis,
takes the form
$M = \left(
\begin{array}{cc}
M_{\nu_L} & m^D_\nu\\
m_\nu^{D^{T}} & M_{\nu_R}
\end{array}
\right)$, where
the submatrix $M_{\nu_L}$ describes the masses
arising in the left-handed sector,
$m_\nu^D$ is the usual Dirac mass matrix
and $M_{\nu_R}$ 
contains the entries in the right-handed isosinglet sector.

A natural question that arises is why neutrino
masses are smaller that the rest of the
fermion masses in the theory. 
This can be explained by the see-saw mechanism \cite{seesaw}.
Suppose that $M_{\nu_L}$ is zero to start with, which is
what happens in the absence of weak-isospin 1 Higgs fields.
Still, a naturally small effective Majorana mass for the light
neutrinos (predominantly $\nu _{L}$) can be generated 
by mixing with the heavy states
(predominantly $\nu _{R})$ of mass $M_{\nu_R}$. 
Indeed, since the Majorana
masses for the right-handed neutrinos
are invariant under the Standard Model gauge group and do not
require a stage of electroweak breaking to generate them,
$M_{\nu_R} \gg M_W$; the Dirac mass matrix
on the other hand  is expected to be similar to the up-quark 
mass matrix
(since neutrinos and up-quarks couple to the same Higgs)
and therefore has similar magnitude. In this case, 
the light eigenvalues of the matrix
$ M = \left( \begin{array}{cc}
0 & m^D_\nu\\
m_\nu^{D^{T}} & M_{\nu_R}
\end{array}
\right)$ are contained in
\begin{eqnarray}
m_{light} \simeq \frac{(m_\nu^{D})^{2}}{M_{\nu_R}}  \nonumber
\end{eqnarray}
and are naturally suppressed.

Then, the neutrino data clearly constrains the possible 
mass scales of the problem.
The mass of the heavier neutrino
is given by  
$\frac{ (m_\nu^D)_{33}^2}{M_{N_3}}$,
where $M_{N_i}$ are the eigenvalues of 
$M_{\nu_R}$.
For a scale ${\cal O} (200 ~ {\rm GeV})$,
solutions of the type (a) (that is light neutrinos of almost equal mass),
require 
\bea
M_{N_3}  \approx {\cal O}({\rm a ~few ~times ~10^{13}} ~ {\rm GeV})
\nonumber 
\eea
Given that the Dirac neutrino couplings  are
expected to have large
hierarchies, we conclude that
in order to obtain three almost degenerate neutrinos, a
large hierarchy
in the heavy Majorana sector is also required.
On the other hand, solutions of the type (b),
with large light neutrino hierarchies
require 
\bea
M_{N_3} \approx O({\rm a ~few ~times ~10^{14}-10^{15} ~{\rm GeV}})
\nonumber 
\eea
The suppression of
$m_{\nu_2}$ with respect to $m_{\nu_3}$ 
can again be obtained either from the Yukawa couplings,
or from the heavy Majorana mass hierarchies:
for $M_{N_2} \approx M_{N_3}$ the relevant squared
Yukawa couplings should have a ratio 1:10.
However, for $M_{N_2} < M_{N_3}$ the ratio
of the relevant squared Yukawa couplings
has to be larger. The same is true for the 
suppression of
 $m_{1}$ with respect to $m_{2}$.
Here, however, the data offer no information on
how large  $m_{1}/m_{2}$ can be.

What about neutrino mixing? This
may occur either purely from the neutrino
sector of the theory, or
by the charged-lepton mixing. Indeed, 
in complete analogy to the quark
currents the leptonic mixing matrix is \cite{mns}
\bea
V_{MNS} =
V_{\ell}V_{\nu}^{\dagger}
\eea
where $V_{\ell}$ diagonalizes the charged-lepton mass matrix, while
$V_{\nu}$ diagonalizes the light neutrino mass matrix.

\section{ Phenomenological Textures}

Let us now try to understand in more detail the
neutrino mass structure that may account for the
various deficits. Initially we will focus on the neutrino
sector of the theory, and in subsequent sections
we will discuss  the possibility that 
the large lepton mixing arises almost
entirely due to the structure of the charged-lepton 
sector \cite{LLR,MG,BPW,Corp}.

We start with
the light-neutrino mass matrix, which 
may be written as
\begin{equation}
m_{eff}=m^D_{\nu}\cdot (M_{\nu_R})^{-1}\cdot m^{D^{\normalsize T}}_{\nu}
\label{eq:meff}
\end{equation}
To identify which mass patterns
may fulfil the phenomenological
requirements outlined in the
previous section, we concentrate initially
on the $2 \times 2$ mass submatrix for the second and
third generations. Then one can write:
\bea
m_{eff}^{-1} = V_{\nu} m_{eff}^{-1\ diag} V_{\nu}^T,
~~~
m^{-1\ diag}_{eff} =
\left (
\begin{array}{ccc}
\frac{1}{m_{2}} & 0 \\
 0 & \frac{1}{m_{3}}
\end{array}
\right) \nonumber 
\eea
where we are
going to explore large (2--3) mixing.
Parametrizing the $2 \times 2$ mixing matrix by
$V_{\nu} = \left
(\begin{array}{ccc}
 c_{23} & -s_{23} \\
 s_{23} & c_{23}
\end{array}
\right)$, 
$m^{-1}_{eff}$ takes the form
\bea
m_{eff}^{-1} & = & \frac 1{m_2m_3} \left (
\begin{array}{cc}
{c^2_{23}}{m_3}+{s_{23}^2}{m_2} &
c_{23}s_{23}({m_3}-{m_2})
\\
c_{23}s_{23}({m_3}-{m_2})
& {c_{23}^2}{m_2}+{s^2_{23}}{m_3}
\end{array}
\right) \nonumber \\
& \equiv & 
d \left (
\begin{array}{cc}
 b/d & 1 \\
 1 & c/d
\end{array}
\right)
\label{eq:form}
\eea

The mass eigenvalues $m_{2,3}$ are given by
\begin{equation}
m_{2,3}  = 
\frac{2}{
b+c \pm 
\sqrt{(b-c)^2+4d^2}} 
\label{mevalues}
\end{equation}
while the
$\nu_{\mu}-\nu_{\tau}$ mixing angle 
\begin{equation}
\sin^2 2\theta_{23}  = 
\left (2 d\frac{ m_2 m_3}{m_3 - m_2} \right ) ^2 =
\frac{4 d^2}{(b-c)^2 + 4 d^2}
\end{equation}
Maximal mixing: $\sin^2 2\theta_{23} \approx 1,\;
\theta_{23} \approx \pi/4$ is obtained 
whenever $|b-c| \ll |d|$. 
Concerning the mass hierarchies, one sees the
following: If the diagonal or the off-diagonal entries dominate,
this results in small hierarchies. On the other hand, if 
all entries are of same order, 
large hierarchies are generated (0-determinant solutions
\cite{guid}).

Having commented on the possible structure
of $m_{eff}$, the
next question is: From what forms of Dirac and heavy
Majorana mass structures may we obtain the
desired $m_{eff}$?
The form of the heavy Majorana mass
matrix $M_{\nu_R}$ 
may easily be found from 
$M_{\nu_R} 
= m_{\nu}^{D^{\normalsize T}}\cdot m_{eff}^{-1}\cdot m_{\nu}^D$
once the neutrino Dirac mass matrix has been
specified. It is clear that if the 
neutrino Dirac mass matrix is diagonal,
one particular solution is 
\beq
M_{\nu_R} \propto (M_{\nu_R} )^{-1} \propto m_{eff} \propto
\left (
\begin{array}{cc}
0 & 1 \\
1 & 0 
\end{array}
\right )
\label{bel}
\eeq
which we discussed in detail in 
\cite{vergados}.

Of course, as the Dirac mass matrix changes,
different forms of $M_{\nu_R}$ are required in order
to obtain the desired form of $m_{eff}$.
This is exemplified in
Table 1, where we show the
textures that lead to 
$m_{eff}$ as given in (\ref{bel})
for three different mixing parameters
in the Dirac mass matrix \cite{ELLN2}.

\vspace*{0.2 cm}

\begin{center}
\begin{small}
\begin{tabular}{|c|c|c|c|} \hline
$m_{\nu}^{D}$ & $(m_\nu^D)^{diag}$ & 
$M_{\nu_R}$  & $M_{\nu_R}^{diag}$  
\\ \hline \hline 
$\left(
\begin{array}{cc}
\la & \la^2 \\
 \la^2 & 1
\end{array}
\right)$ &
$\left(
\begin{array}{cc}
\la & 0\\
0 & 1
\end{array}
\right)$  &
$  \left(
\begin{array}{ccc}
2 \la^2  & 1 \\
1 & 2 \la 
\end{array}
\right)$ &
$ \left(
\begin{array}{cc}
-1 & 0\\
0 & 1
\end{array}
\right)$  
\\ 
\hline
$\left(
\begin{array}{cc}
\la^3 & \la^2 \\
\la^2 & 1
\end{array}
\right)$ &
$\left(
\begin{array}{cc}
\la^3 & 0 \\
0 & 1
\end{array}
\right)$ &
$  \left(
\begin{array}{ccc}
2\la^3  & \la \\
\la & 2 
\end{array}
\right)$ &
$  \left(
\begin{array}{ccc}
-\la^2/2  & 0 \\
0 & 2 
\end{array}
\right)$
\\ 
\hline
$\left(
\begin{array}{cc}
\la & 1 \\
 1 & \la
\end{array}
\right)$ &
$\left(
\begin{array}{cc}
-1 & 0 \\
 0 & 1
\end{array}
\right)$ &
$ \left(
\begin{array}{ccc}
2 \la  & 1 \\
1 & 2 \la 
\end{array}
\right)$ &
$ \left(
\begin{array}{ccc}
-1  & 0 \\
0 & 1  
\end{array}
\right)$
\\ \hline
\end{tabular}
\end{small}
\end{center}

\vspace*{0.2 cm}

{\small Table 1: {\it Approximate forms for some of the
basic structures of symmetric
textures, keeping the dominant contributions.}}%

\vspace*{0.4 cm} 

The above,
can be described in a more generic way:
for simplicity, we consider the case of a symmetric Dirac mass matrix
with mixing angle $\vartheta$. We define
$\phi$ to be the mixing angle in the heavy Majorana neutrino mass matrix,
and denote by $\theta$ the resulting mixing angle in 
the light-neutrino mass matrix $m_{eff}$
(where from now on we drop the sub-indices that
refer to the (2--3) sector).
$M_{\nu_R}$ can be parametrised as \bea
M_{\nu_R} = \left (
\begin{array}{cc}
M_2\cos^2\phi + M_3 \sin^2\phi &
(M_2-M_3)\cos\phi \sin\phi \\
(M_2-M_3)\cos\phi \sin\phi &
M_3\cos^2\phi + M_2 \sin^2\phi 
\end{array}
\right ) \nonumber 
\label{relate}
\eea
where the mixing angle 
is given by \cite{ELLN2}
\bea
\tan 2 \phi = 
\frac{ 
\sin (4\vartheta-2\theta)+r^2\sin 2\theta - 2 r R \sin 2 \vartheta
}
{
\cos (4\vartheta-2\theta)+r^2 \cos 2\theta - 2 r R \cos2 \vartheta
} 
\label{EQMIX}
\eea
Here, $M_3$ and $M_2$ are the eigenvalues of the
heavy Majorana mass matrix,
$R \equiv
(m_2+m_3)/(m_3-m_2)$ with
$m_i$ being the eigenvalues of the light-neutrino mass matrix, and
$r \equiv
(m^D_2+m^D_3)/(m^D_3-m^D_2)$, with the $m^D_i$ being the
eigenvalues of the Dirac mass matrix.
Eq. (\ref{EQMIX}) relates the mass and mixing parameters
of the various neutrino sectors.

The $2 \times 2$ description may be a good approximation
in the limit
where the solar neutrino problem is
resolved by a small mixing angle.
However, this need not be the case, and one
should consider the $3\times 3$ mixing
problem \footnote{
Here we follow the discussion of \cite{ELLN2},
however there are many recent papers 
containing such types of textures \cite{recent}. 
}.
The generic form of a $3 \times 3$ mixing-angle matrix
(ignoring phases) is
\bea
V_{3 \times 3} = \left
(\begin{array}{ccc}
c_{12} & -s_{12} c_{13} & -s_{12} s_{13} \\
s_{12} c_{23} & c_{12}c_{23}c_{13}+s_{23}s_{13} & c_{12}c_{23}s_{13}-s_{23}c_{13} \\
s_{12}s_{23} & c_{12}s_{23}c_{13}-c_{23}s_{13} & c_{12}s_{23}s_{13} + c_{23}c_{13}
\end{array}
\right) \nonumber 
\eea
where $s_{ij},c_{ij}$ stand for 
$\sin\theta_{ij}$ and $\cos\theta_{ij}$,
respectively. Investigating the 
possible hierarchies
within $m_{eff}$ is then straightforward,
since it is given by
$m_{eff} = V_{3 \times 3}.m_{eff}^{diag}. V_{3 \times 3}^{\dagger}$.
When specific limits are
considered, simple expressions for
$m_{eff}$ can be derived. For example, for 
maximal $\theta_{12},\theta_{23}$ mixing
and  $\theta_{13} \sim 0$, 
in the limit $m_3 \gg m_2 \gg m_1$, one has:
\bea
m_{eff} & = & {m_3 \over 2} \left(
\begin{array}{ccc}
0 & 0 & 0 \\
0 & 1 & -1 \\
0 & -1 & 1
\end{array}
\right) + {m_2 \over 2} \left(
\begin{array}{ccc}
1 & -{1 \over \sqrt{2}} & -{1 \over \sqrt{2}} \\
-{1 \over \sqrt{2}} & {1 \over 2} & {1 \over 2} \\
-{1 \over \sqrt{2}} & {1 \over 2} & {1 \over 2}
\end{array}
\right) \nonumber \\
& + & {m_1 \over 2} \left(
\begin{array}{ccc}
1 & {1 \over \sqrt{2}} & {1 \over \sqrt{2}} \\
{1 \over \sqrt{2}} & {1 \over 2} & {1 \over 2} \\
{1 \over \sqrt{2}} & {1 \over 2} & {1 \over 2}
\end{array}
\right)
\label{expansion}
\eea
Analogous expressions are obtained in the case where the
neutrino masses exhibit some degeneracy.

What about the structure of the Dirac and 
heavy Majorana matrices that generate  viable 
$m_{eff}$'s in this case? In view of the many parameters,
at this stage we look at some limiting cases
for symmetric Dirac mass matrices
(and subsequently we will examine solutions
in models with flavour symmetries, 
including also asymmetric textures).
It is convenient to
parametrise the output  in terms of the hierarchy factors
$x \equiv m_1 / m_3, y \equiv m_2 / m_3$ for 
the ratios of eigenvalues of $m_{eff}$
and $\lambda_1 \equiv m_{\nu_1}^D / m_{\nu_3}^D,
\lambda_2 \equiv m_{\nu_2}^D / m_{\nu_3}^D$ for
the ratios of eigenvalues of the neutrino Dirac mass
matrix $m_{\nu}^D$.

(A) We can distinguish two cases for
the structure of the heavy Majorana matrix:
The first is that of {\it matched mixing}, which
occurs when we have one large mixing angle
in the (2--3) sector of $m_{eff}$
and there is no large mixing
in other sectors of either
the light Majorana or the Dirac matrices. In this case, the
problem is equivalent to the $2\times 2$ case considered previously.
In the particular cases that $y= m_2 / m_3 = -1$ and
$x \ll y \ll 1$, one obtains the  textures
\bea
M_{\nu_R} \propto  \left
(\begin{array}{ccc}
\l_1^2\over  x & 0 & 0 \\
0 & 0 & \l_2 \\
0 & \l_2 & 0
\end{array}
\right) ~~ {\rm and} ~~ 
M_{\nu_R} \propto  \left
(\begin{array}{ccc}
\l_1^2\over  x & 0 & 0 \\
0 & \l_2^2\over  2 y & \l_2\over   2 y \\
0 & \l_2\over  2 y & 1\over  2 y
\end{array}
\right) \nonumber 
\eea
respectively, which indicate the decoupling of the
light sector.

(B) A different structure arises when 
(i) there is more than one
mixing angle in $m_{eff}$ and/or
(ii) there is a large
Dirac mixing angle that involves
different generations from those of the light Majorana matrix.
This happens, for example when the atmospheric problem is solved
by $\nu_{\mu}\ra \nu_{\tau}$ oscillations, whilst the Dirac mass
matrix is related to the quark mass matrix, with Cabibbo mixing
between the first and second
generations. The structure of the Majorana matrix
becomes more complicated for this {\it mismatched mixing}.
It is interesting to note that,
for an almost-diagonal
Dirac mass matrix and  large Dirac hierarchies 
(and in particular 
$\lambda_1 \ll \lambda_2$),
the light entry of the
heavy Majorana mass matrix again
effectively decouples from the
heavier ones \cite{ELLN2}.
This is no longer true, however, if
the (12) mixing angle in the
Dirac mass matrix increases.
For example, for maximal (1-2) Dirac mixing 
(which is plausible, as we discuss below),
two large mixing angles in $m_{eff}$ ($\theta_{23}$ and
$\theta_{12}$)
and
large hierarchies
$y \ll 1$ and 
$\l_2^2 x \gg \l_1^2 y$,
one has 
\bea
M_{\nu_R} \propto  \frac{1}{2y}
\times  \left
  (\begin{array}{ccc}
\l_2^2\over  2 & \l_2^2\over  2 & -\l_2\over  \sqrt{2}  \\ 
\l_2^2\over  2 & \l_2^2\over  2 & -\l_2\over  \sqrt{2}  \\ 
-\l_2\over  \sqrt{2}  & -\l_2\over  \sqrt{2}  & 1
  \end{array}
\right) 
\eea
while for intermediate (1-2) Dirac mixing the
effect lies between the two limiting cases that we discussed.
Since there is a vast number of possibilities for
the origin of proper mixings, we will investigate the
type of constraints that one may obtain from flavour symmetries.
Before doing so, however, we will discuss the
effect of the running of couplings to the neutrino
textures.

\section{ Renormalisation group effects}

In the presence of neutrino masses,
the running of the various couplings
from the unification scale
down to low energies is modified.
From $M_{GUT}$ to $M_N$, one  must include
 radiative corrections  from  $\nu_R$ neutrinos, while
below $M_N$, the $\nu_R$'s decouple from the 
spectrum and an effective  see-saw mechanism is operative.
It actually turns out that the renormalisation group
effects (which may be demonstrated by simple semi-analytic
expressions) give important information on
the structure of the neutrino textures.
To see this,  let us start 
with the small $\tan\beta$ regime of a supersymmetric
theory. In this case, only the top and the Dirac-type 
neutrino Yukawa coupling, $\lambda_N$,  may be large at the GUT scale
(approximate equality of $\lambda_t$ and $\lambda_{N}$ 
arises naturally in many Grand Unified Models;
for a smaller $\lambda_{N}$ at $M_{GUT}$, the effects 
that we will describe reduce accordingly).
In a diagonal basis,
the renormalization
group  equations for the Yukawa couplings at the one-loop level
 can be written  as follows \cite{VB}:
 \bea
 16\pi^2 \frac{d}{dt} \lambda_t&=
 & \left(
 6 \lambda_t^2  + \lambda_N^2
   - G_U\right)  \lambda_t \nonumber \\
 16\pi^2 \frac{d}{dt} \lambda_N&=& \left(
  4\lambda_N^2  + 3 \lambda_t^2
   - G_N \right) \lambda_N \nonumber   \\
 16\pi^2 \frac{d}{dt} \lambda_b &=
 & \left(\lambda_t^2 - G_D \right) \lambda_b \nonumber \\
 16\pi^2 \frac{d}{dt} \lambda_{\tau}&=&\left( \lambda_N^2
  - G_E \right) \lambda_{\tau}
 \label{eq:rg4}
 \eea
 Here, $\lambda_\alpha$, $\alpha=U,D,E,N$, represent the
 $3 \times 3$ Yukawa matrices for the up and down quarks, charged
lepton and Dirac neutrinos,
and $G_{\alpha}= \sum_{i=1}^3c_{\alpha}^ig_i(t)^2$ are
functions that depend on the  gauge couplings with the
coefficients $c_{\alpha}^i$'s as in \cite{VB}.
Let us denote by $\lambda_G$ the top and neutrino Yukawa
couplings,
and by $\lambda_{b_0},{\lambda_{\tau_0}}$
the $b$ and $\tau$ couplings
at the  unification scale. Then 
  \begin{eqnarray}
 \lambda_t(t)&=&\gamma_U(t)\lambda_G\xi_t^6\xi_N ~~~~~ \lambda_N(t)=\gamma_N(t)\lambda_G\xi_t^3\xi_N^4\\
 \lambda_b(t)&=&\gamma_D(t)\lambda_{b_0}\xi_t ~~~~~~~~~ \lambda_{\tau}(t)=\gamma_E(t)\lambda_{\tau_0}\xi_N
 \end{eqnarray}
 where $\gamma_\alpha(t)$ and
 $\xi_{i}$ depend purely on
 gauge coupling constants and Yukawa couplings respectively:
 \bea
 \gamma_\alpha(t)&=&  \exp\left ({\frac{1}{16\pi^2}\int_{t_0}^t
  G_\alpha(t) \,dt} \right ) = \prod_{j=1}^3 \left( \frac{\alpha_{j,0}}{\alpha_j}
 \right)^{c_\alpha^j/2b_j} \nonumber \\
 \xi_i&=& \exp
\left ({\frac{1}{16\pi^2}\int_{t_0}^t \lambda^2_{i}dt} \right )
 \eea
 One then finds that
 \begin{equation}
 \lambda_{b}(t_N)=\rho
 \xi_t\frac{\gamma_D}{\gamma_E}\lambda_{\tau}(t_N), ~~~
\rho=\frac{\lambda_{b_0}}{\lambda_{\tau_0}\xi_N}
 \end{equation}
For $b$--$\tau$ unification at $M_{GUT}$, 
  $\lambda_{\tau_0} =\lambda_{b_0}$. In the absence
 of the right-handed neutrino $\xi_N \equiv 1$, thus
 $\rho =1 $ and $m_b$  at low
 energies is correctly predicted.
 In the presence of $\nu_R$, however,  $\lambda_{\tau_0}
 =\lambda_{b_0}$ at the GUT scale implies that 
$\rho \neq 1$  (since $\xi_N<1$). 
To restore $\rho$ to unity, a deviation from bottom--tau unification
is required. For example,
for $M_N \approx 10^{13}~{\rm GeV}$ and 
$\lambda_G \ge 1$, it turns out that
 $\xi(t_N)\approx 0.89$. This corresponds to an
approximate  $10\%$ deviation of the $\tau$--$b$ equality at the
GUT scale, in agreement with the numerical results.

For large $\tan\beta$, one expects
large corrections to $m_b$ \cite{Hal,CW}.
Even ignoring these corrections, the effect
of the heavy neutrino scale is much smaller, since now
the bottom Yukawa coupling also runs to a fixed point
\cite{CW}.
For large $\tan\beta$, and $\lambda_b\approx \lambda_{t}$,
the product and ratio of the top and bottom
couplings can be simply expressed as
$ \lambda_t \lambda_b \approx
\frac{{8\pi^2}\gamma_Q \gamma_D}{{7}\int \gamma_Q^2 d\,t},
~\frac{\lambda_t^2}{\lambda_b^2}\approx \frac{\gamma_Q^2}{\gamma_D^2}$ 
 \cite{FLL},
indicating that one gets an approximate,
model-independent  prediction for both couplings
at the low-energy scale.

Given these results, it is natural to ask if
models with $b$--$\tau$
equality and large neutrino Yukawa couplings
at $M_{GUT}$ may be consistent with the 
required neutrino masses in the small $\tan\beta$ regime.
To answer this, we need to remember that
the $b$--$\tau$ equality at the GUT scale refers to the
$(3,3)$ entries of the charged lepton and
down quark mass matrices, while the detailed structure of the
mass matrices is not predicted by the Grand Unified
Group itself.
It is then possible to assume mass textures,
such that, after the diagonalisation at   the
GUT scale, the $(m^{diag}_E)_{33}$ and $(m^{diag}_D)_{33}$
entries are no-longer equal \cite{LLR}.

To quantify the effect, we worked with a simple $2\times 2 $ example.
Let us assume a diagonal form of $m_D$ at the GUT scale,
 $m_D^0 = diagonal (c m_0,m_0)$, while the corresponding
 entries of charged-lepton mass matrix have the form
$ m_{E}^0 =
 \left (
 \begin{array}{cc}
 d & \tilde{\epsilon} \\
 \tilde{\epsilon} &  1
 \end{array}
 \right) m_0 $,
 ensuring that at the GUT scale
 $(m_D^0)_{33}= (m_E^0)_{33}$. At low energies,
the eigenmasses are obtained by
diagonalising the renormalised Yukawa matrices. This is equivalent to
 diagonalise the quark and charged lepton Yukawa matrices
 at the GUT scale and evolve the eigenstates and
 the mixing angles separately. Since $m_D^0$ has been chosen diagonal,
$ m_s=c \gamma_D m_0$ and
$ m_{b} =  \gamma_D  m_0 \xi_t$,
 with $m_0 = \lambda_{b_0} \frac{\upsilon}{\sqrt{2}}cos\beta$.

 To find the charged-lepton mass eigenstates we need first to
 diagonalise $m_E^0$ at $M_{GUT}$. One finds that
 \bea
 d= \left (\frac{m_{\tau}^0 - m_{\mu}^0}{m_0}-1 \right ), \; \; \;
 \tilde{\epsilon}^2
  = \left ( \frac{m_{\mu}^0}{m_0}+1 
\right )  \left ( \frac{m_{\tau}^0}{m_0}-1 \right )
 \eea
 The evolution of
 the $\tau$--eigenstate down to low energies (in the presence
of neutrinos) is
 described  by (\ref{eq:rg4}) with
 $m_{\tau_0}=\lambda_{\tau_0}
 \frac{\upsilon}{\sqrt{2}}\cos\beta$.  
Then, it turns out that obtaining the
 correct $m_{\tau}/m_b$ ratio at low energies while preserving 
 $b$-- $\tau$  unification at $M_{GUT}$, 
requires \cite{LLR}
 \bea
 \tilde{\epsilon} = \sqrt{\frac{1}{\xi_N}-1} ,&
 d \approx (\frac{1}{\xi_N}-1) = \tilde{\epsilon}^2
\label{eq:de}
 \eea
and thus a non-trivial $\mu$--$\tau$ mixing, 
which depends only on the scale $M_N$ and the initial
 $\lambda_N$ condition. For example, for  $M_N \approx 10^{13}$,
the mixing in the
charged lepton sector alone has to be
$\sin^2 2 \theta (M_{GUT}) \approx 0.4$.
This mixing may be amplified from the one 
in the light neutrino sector, in a way that the total
mixing reaches  the values required by
Super-Kamiokande.

Below the right-handed Majorana mass scale, 
${\lambda}_N$ decouples and the relevant running
is that of the effective neutrino mass operator:
\bea
    8\p^2 {d\over dt}{ \k} & = & \{-({3\over 5} g_1^2+3 g_2^2)+
{\rm Tr} [3{ \lambda}_U { \lambda}_U^\dagger]\}\k \nonumber \\
& &  +{1\over 2}\{({ \lambda}_E { \lambda}_E^\dagger)\k +
   \k ({ \lambda}_E { \lambda}_E^\dagger)^T\}  
\label{MASSES}
\eea
We already see that large Yukawa terms,
which lower the effective couplings,
have a larger effect on $m_{eff}^{33}$ than
on the other entries.
The running of 
the neutrino mixing angle $\th_{23}$
is given by \cite{Bab,run}
\bea
    16\p^2 {d\over dt}{\sin^2 2\th_{23}} & =& -2\sin^2 2\th_{23}
        (1-\sin^2 2\th_{23}) \nonumber \\
& &  (\lambda_{\tau}^2-\lambda_{\mu}^2){\k^{33}+\k^{22}\over
\k^{33}-\k^{22}}  
\label{MIXING}
\eea
This already indicates that sin$^2 2\th_{23}$ 
is significantly changing from the
GUT scale to low energies
(i) if $\lambda_{\tau}$ is large, and (ii)
if the diagonal entries of $m_{eff}$ are 
close in magnitude \cite{Bab}.
To quantify this statement analytically,
we integrate the differential
equations for the diagonal elements
of  the effective neutrino mass matrix \cite{ELLN2}.
This yields the result  
\bea
\frac{m_{eff}^{33}}{m_{eff}^{22}} = I_{\tau}\cdot
\frac{m_{eff,0}^{33}}{m_{eff,0}^{22}}
\label{bb}
\eea
where
\bea
    I_{\tau} &=& \exp \left [ \frac 1{8\pi^2}\int_{t_0}^t \lambda_{\tau}^2 dt
\right ]
\eea
and $ m_{eff,0}^{33(22)}$ is the initial
condition, defined at the stage when $\lambda_N$ decouples from
the renormalisation-group equations. For
simplicity of presentation, we assume here that
$M_N \approx M_{GUT}$.

Eq.(\ref{bb}) leads to the result
\bea
   \frac{m_{eff}^{33}+m_{eff}^{22}}{m_{eff}^{33}-m_{eff}^{22}}& =&
    \frac{m_{eff,0}^{33} I_{\tau} +m_{eff,0}^{22}}
{m_{eff,0}^{33} I_{\tau}-m_{eff,0}^{22}} \equiv  f(I_{\tau})
\eea
We can then convert
the one-loop evolution equation (\ref{MIXING}) for sin$^2\theta$ to a 
differential equation for $T = \tan^22\theta$,
and its solution is \cite{ELLN2}
\bea
    \tan^2 2\theta &=& \tan^2 2\theta_0 I_2(\lambda_{\tau})
\label{tan}
\eea
with 
\bea
I_2(\lambda_{\tau}) =  exp\left\{ -\frac{1}{8\pi^2} \int_{t_0}^t
                  \lambda_{\tau}^2 f(I_{\tau})\right\} 
\eea

Let us see what conclusions we can draw
from these formulae, without doing any numerical
analysis: we first see that
$m_{eff}^{33}$ decreases
more rapidly than
$m_{eff}^{22}$, due to the effect of the
$\tau$ Yukawa coupling. The effect is much more significant
for large $\tan\beta$ (where
$\lambda_\tau$ is large), while for small $\tan\beta$
the effects are negligible.
In the former case, if one starts with
 $m_{eff}^{22} < m_{eff}^{33}$ and 
$m_{eff}^{22}$, $m_{eff}^{33}$ 
relatively close in magnitude
the expectation is that at a given scale
they may become equal, in which case the mixing angle is {\em maximal}.
The larger $\lambda_{\tau}^0$,  the earlier the entries
may become equal.
The exact scale where the mixing angle is maximal
is given by the relation
\bea
I_{\tau} = \frac{m_{eff,0}^{22}}{m_{eff,0}^{33}}
\eea
After reaching the maximal angle at some
intermediate scale, the running of
$\lambda_{\tau}$ results in
\bea
{m_{eff,0}^{33}}  < {m_{eff,0}^{22}} 
\nonumber
\eea
This changes the sign of
$f(I_{\tau})$ and results in a rather rapid decrease of
the mixing. 
In order, therefore, for a texture of this type to be
viable, there needs to be a balance between
the magnitudes of $\lambda_{\tau}$ and
$m_{eff}^{33}-m_{eff}^{22}$ at the GUT
scale. For a proper choice of parameters, a
small mixing at $M_{GUT}$ may be converted to
maximal mixing at low energies. 
If, however, the splitting is small and
the coupling large, then the maximal value 
for the mixing will be obtained too early
to survive at low energies. 
These considerations may impose strong constraints on 
certain types of phenomenological textures.

\section{Fermion masses from \mbox{\boldmath{$U(1)$}}
and GUT symmetries}

The fact that the fermion mass matrices exhibit a hierarchical
structure suggests that they are generated by 
an underlying family symmetry \cite{FN,textures,neutr,IR}. 
Here we consider the simplest possibility, where the
flavour-symmetry is abelian, and we denote 
the  charges of the Standard model 
fields under the symmetry as appear in Table 2.

\vspace*{0.2 cm}

\begin{center}
\begin{tabular}{|c|cccccccc|}
\hline
& $Q_{i}$ & $u_{i}^{c}$ & $d_{i}^{c}$ & $L_{i}$ & $e_{i}^{c}$ & $\nu
_{i}^{c} $ & $H_{2}$ & $H_{1}$ \\ \hline
$U(1)$ & $\alpha _{i}$ & $\beta _{i}$ & $\gamma _{i}$ & $b_{i}$ & $%
c_{i} $ & $d_{i}$ & $-\alpha _{3}-\beta _{3}$ & $-\alpha _{3}-\gamma _{3}$
\\ \hline
\end{tabular}
\end{center}

\vspace*{0.2 cm}

{\small Table 2: {\it $U(1)$ charges of the various
fields,
where $i$ stands for a generation index.}}

\vspace*{0.4 cm}

The Higgs charges are chosen so that the
terms $f_3 f^c_3 H$ (where $f$ stands for a fermion
and $H$ denotes
$H_{1}$ or $H_{2}$) have zero charge.
Then, only the (3,3) element 
of the associated mass matrix will be non-zero. 
The remaining entries are generated when the
$U(1)$ symmetry is spontaneously broken, via 
standard model singlet fields,
$\theta,\; \bar{\theta}$, with $U(1)$ charge $-1$, $+1$
respectively and equal vevs (vacuum expectation values).
The suppression factor for each entry depends on the family charge:
the higher the net $U(1)$ charge of a term $f_i f^c_j H$,
the higher the power $n$ in the non-renormalisable term
$f_i f^c_j H \left ( \frac{\theta}{M} \right)^n $
that has zero charge.
For example, if only 
the 2--3 and 3--2 elements 
of the matrix are allowed by the symmetry at order $\epsilon \equiv
\theta /M$,
one has the following hierarchy of masses:
\begin{equation}
{\cal M}\sim \left( 
\begin{array}{ccc}
0 & 0 & 0 \\ 
0 & 0 & 0 \\ 
0 & 0 & 1
\end{array}
\right) \rightarrow \left( 
\begin{array}{ccc}
0 & 0 & 0 \\ 
0 & 0 & \epsilon \\ 
0 & \epsilon & 1
\end{array}
\right)
\end{equation}
where $M$ is an intermediate mass 
scale,  determined by the mechanism that generates 
the non-renormalisable terms. A common approach
communicates symmetry breaking via an extension of the ``see-saw'' mechanism,
mixing light to heavy states  and is known as the 
Froggatt--Nielsen mechanism \cite{FN}.

An interesting question that immediately arises is whether
realistic fermion mass structures are consistent
with the constraints on an Abelian family symmetry 
in GUT-embedded solutions and, if yes, which 
GUT schemes would be favoured.  We worked along these lines, 
for a variety of GUT models and under the
assumption of large $\tan\beta$ (which 
implies equal charges for $H_1$ and $H_2$).
As a result, we observed the following \cite{MG}:

{\bf (A)~} In an \mbox{\boldmath{$SO(10)$}} GUT, all quark 
and lepton charges for the left- and 
right-handed fields of a given family are the same,
leading to left-right-symmetric mass matrices with similar structure
for all fermions. Since
the down quarks and charged leptons couple to
the same Higgs,  one has the prediction
$ V_{\mu {\tau }}\approx V_{cb} $,
which  may only be reconciled with observations
either with the help of coefficients, or by
assuming that the heavy fields
responsible for the Froggatt--Nielsen mixing 
have restricted $U(1)$ family
charges.

{\bf (B)~} In \mbox{\boldmath{$SU(5)$}},
the field structure is
$(Q,u^{c},e^{c})_{i} \in {\tt 10}$ of $SU(5)$
and
$(L,d^{c})_i  \in { \tt \overline{5}}$, implying the following:
(i) the 
up-quark mass matrix is symmetric and 
(ii) the charged-lepton mass matrix is the transpose of the 
down-quark mass
matrix, thus relating the left-lepton with the
right-quark mixing. This explains how the large mixing angle  that
is observed in atmospheric
neutrinos can  be consistent with the small $V_{CKM}$ mixing,
without any tuning. On the other hand, 
obtaining the correct $V_{CKM}^{12,21}$ 
(arising from the down-quark sector),
inevitably leads to a larger $m_{up}$ than 
indicated by the data. The abelian
symmetry alone may not guarantee the smallness
of $m_{up}$ without introducing large
coefficients or cancellations. However it has been
proposed  that such a small term may in principle be generated 
by alternative means \cite{guid2}.

{\bf (C)~} In the case of the 
{\bf flipped}-\mbox{\boldmath{$SU(5)$}}, the 
fields $Q_{i},d_{i}^{c}$ and $\nu
_{i}^{c}$ belong to a 
{\tt $10$} of $SU(5)$, while $u_{i}^{c}$ and $%
L_{i}$ belong to a ${\tt \overline{5}}$. Finally,
the $e_{i}^{c}$ fields belong to
singlet representations of $SU(5)$. This 
assignment implies symmetric down-quark mass matrices.
The structure of the up-quark mass matrix will depend on the charges of the
right-handed quarks. However, as these are the same with the charges of the
left-handed leptons, the mass matrix will be constrained by the need to
generate large mixing for atmospheric neutrinos. 
In this model, it turns out  that the
contribution from the up-quark sector to $V_{cb}$ is negligible \cite{MG}
and therefore $V_{cb}\simeq \sqrt{%
m_{s}/m_{b}.}$ This is too large and requires a significant coefficient
adjustment.
However, as we will discuss later, the string-embedded flipped
$SU(5)$ model, due to its additional
(although highly constrained) structure, works in a nice way.

{\bf (D)~} Under 
\mbox{\boldmath{$SU(3)_{c}\times SU(3)_{L}\times SU(3)_{R}$}}
the left- and right-handed quarks belong to a $(3,3,1)$ and $(\bar{3},1,\bar{3}%
)$ respectively and thus their $U(1)$ charges are not related. On the other
hand the left-handed and (charge conjugate) right-handed leptons belong to
the same $(1,3,\bar{3})$ representation and hence must have the same $U(1)$
charge. Thus, the lepton mass matrices have to be symmetric.
This freedom allows us to construct fully realistic mass matrices \cite{MG}.
Since the quark mass matrices are asymmetric
(with different expansion parameters but  similar
structure for up- and down-quarks), it is
straightforward to chose $U(1)$ charges, such that all quark hierarchies
are fulfilled. This choice, does not impose any
constraints on the lepton charges. Concerning the latest,
the choice of charges 
\begin{eqnarray}
b_{i}=c_{i}=d_{i} &=& \left (-\frac{7}{2},\frac{1}{2},0 \right ) 
\nonumber  \\
b_{i}=c_{i}=d_{i} &=& \left (\frac{5}{2},\frac{1}{2},0 \right ) 
\nonumber   \\
b_{i}=c_{i}=d_{i} &=&(3,0,0)
\label{char}
\end{eqnarray}
leads to the three possible charged-lepton matrices :
\bea
M_{\ell }  \propto  \left( 
\begin{array}{ccc}
\bar{\epsilon}^{7} & \bar{\epsilon}^{3} & \bar{\epsilon}^{7/2} \\ 
\bar{\epsilon}^{3} & \bar{\epsilon} & \bar{\epsilon}^{1/2} \\ 
\bar{\epsilon}^{7/2} & \bar{\epsilon}^{1/2} & 1
\end{array}
\right) ,
M_{\ell } &  \propto & \left( 
\begin{array}{ccc}
\bar{\epsilon}^{5} & \bar{\epsilon}^{3} & \bar{\epsilon}^{5/2} \\ 
\bar{\epsilon}^{3} & \bar{\epsilon} & \bar{\epsilon}^{1/2} \\ 
\bar{\epsilon}^{5/2} & \bar{\epsilon}^{1/2} & 1
\end{array}
\right)  \nonumber \\
M_{\ell } & \propto & \left( 
\begin{array}{ccc}
\bar{\epsilon}^{6} & \bar{\epsilon}^{3} & \bar{\epsilon}^{3} \\ 
\bar{\epsilon}^{3} & 1 & 1 \\ 
\bar{\epsilon}^{3} & 1 & 1
\end{array}
\right) 
\label{LEPTONS} 
\eea
We see that the third matrix leads to maximal mixing, however, it requires an
accurate cancellation in the (2,3) sector in order to get the correct $%
m_{\mu }/m_{\tau }$. On the other hand, the other two matrices lead to
natural lepton hierarchies for $\bar{\epsilon} \approx 0.2$ 
and imply large but non-maximal lepton mixing
(which we discuss later in this section).
Since this model has symmetric lepton matrices, 
its predictions for charged lepton
and neutrino masses will be along the lines
of left-right symmetric models,
which we proceed to discuss.

{\bf (E)~} In {\bf Left-Right symmetric models},
the $U(1)$ family charges are strongly constrained because the 
symmetry requires that the $U(1)$ charges of the left- and 
right-handed fields be identical. 
A model for fermion masses in this framework,  has been
proposed in \cite{IR}\footnote{
Note that the hierarchies of lepton(baryon)-number
violating operators as well as of soft-terms
may be discussed in a similar way.
For example, in the framework of the \cite{IR} construction, 
these analyses has been performed in \cite{Rviol} and
\cite{LFV} respectively
.}.
 In this construction,
{\em all } the mass matrices will be symmetric, however,
unlike the minimal $SO(10)$, the quark and lepton mass
matrices need not have the same structure.
The lepton sector is identical to that of
$SU(3)_{c}\times SU(3)_{L}\times SU(3)_{R}$,
thus the charged lepton matrices are as
in (\ref{LEPTONS}).

What about neutrino masses? 
The neutrino Dirac mass is specified to be
of the same type as
for the charged leptons, but with a different
expansion parameter. Indeed,
since neutrinos (charged leptons) and up-type (down-type)  quarks
couple to the same Higgs, they should have the same expansion 
parameter $\epsilon (\bar{\epsilon})$,
where the spread between the up- and down-quark hierarchies
requires $\epsilon \approx  \bar{\epsilon}^2$. Then, 
\bea
m^D_{\nu} \propto \left( 
\begin{array}{ccc}
{\epsilon}^{7} & {\epsilon}^{3} & {\epsilon}^{7/2} \\ 
{\epsilon}^{3} & {\epsilon} & {\epsilon}^{1/2} \\ 
{\epsilon}^{7/2} & {\epsilon}^{1/2} & 1
\end{array}
\right),~
m_{\nu }^D \propto \left( 
\begin{array}{ccc}
{\epsilon}^{5} & {\epsilon}^{3} & {\epsilon}^{5/2} \\ 
{\epsilon}^{3} & {\epsilon} & {\epsilon}^{1/2} \\ 
{\epsilon}^{5/2} & {\epsilon}^{1/2} & 1
\end{array}
\right)  \nonumber
\eea
for the first two choices of charges in 
(\ref{char}) respectively.

Of course the mass structure of neutrinos is more
complicated, due to the heavy Majorana masses of the
right-handed components. These arise
from a term of the form
$\nu_R\nu_R\Sigma$,  where $\Sigma$ is a $SU(3)\times
SU(2)\times U(1)$ invariant Higgs scalar field with $I_W=0$.
Looking in more detail at the origin of neutrino masses in relevance to the
breaking of the left-right symmetry, one
concludes \cite{DLLRS} that the appropriate expansion parameter for the
Majorana  mass matrix is the same as that for the {\it down}
quarks and charged  leptons. Then, the possible choices for the
$\Sigma$  charge
 will give a discrete spectrum of
possible forms for the Majorana mass,
$M_{\nu_R}$ \cite{DLLRS,LLR}.
For example, if $\Sigma$ has the same charge with
the Higgs doublets, the form of the heavy Majorana mass matrix 
will be similar to that of the charged leptons.
For simplicity of presentation, here
we isolate this choice of $\Sigma$ charge and discuss
the set of textures that result
out of the solution with
$b_{i}=c_{i}=d_{i} = \left (\frac{5}{2},\frac{1}{2},0 \right ) $.
For a complete
study of the mass matrices, we refer to
\cite{MG}. For the particular choice made here, we find that:
\bea
m_{eff} = \left (
\begin{array}{ccc}
\bar{\epsilon}^{10} & \bar{\epsilon}^{6} & \bar{\epsilon}^{5} \\ 
\bar{\epsilon}^{6} & \bar{\epsilon}^{2} & \bar{\epsilon} \\ 
\bar{\epsilon}^{5} & \bar{\epsilon} & 1
\end{array}
\right),~
m_{eff}^{diag} = 
 \left( 
\begin{array}{ccc}
\bar{\epsilon}^{15} &  &  \\ 
& \bar{\epsilon}^{3} &  \\ 
&  & 1
\end{array}
\right) \nonumber 
\eea
\bea
V_\ell = \left( 
\begin{array}{ccc}
1 & \bar{\epsilon}^{2} & -\bar{\epsilon}^{5/2} \\ 
-\bar{\epsilon}^{2} & 1 & \bar{\epsilon}^{1/2} \\ 
\bar{\epsilon}^{5/2} & -\bar{\epsilon}^{1/2} & 1
\end{array}
\right), V_\nu = \left( 
\begin{array}{ccc}
1 & \bar{\epsilon}^{4} & -\bar{\epsilon}^{5} \\ 
-\bar{\epsilon}^{4} & 1 & \bar{\epsilon} \\ 
\bar{\epsilon}^{5} & -\bar{\epsilon} & 1
\end{array}
\right) \nonumber 
\eea

One then sees the following:

$\bullet$
Since in these solutions $m^D_\nu$ and 
$M_{\nu_R}$ have a hierarchical structure, so
does $m_{eff}$.

$\bullet$ 
The 2--3 mixing from $V_\nu$ is of ${\cal O}
(\overline{%
\epsilon })$, while the contribution to the 2--3 mixing angle
from the charged-lepton sector 
is of ${\cal O}(\sqrt{\overline{\epsilon }})$
(where $\bar{\epsilon}$ has been fixed to $\approx 0.2$
from the quark and charged lepton hierarchies).
In the case that the two
sources of mixing act constructively,
a total 2--3 mixing with $\sin\theta $ up
to $\sqrt{\overline{\epsilon }}+\overline{\epsilon }\approx 0.7$ is
obtained \footnote{
Here we should stress that, although for illustrative purposes we fixed the form
of $M_{\nu_R}$, the contribution to $V_{MNS}$
from the neutrino sector is not sensitive
to the structure of $M_{\nu_R}$. 
This is indicative of the
fact that the mixing in models
with large hierarchies and non-zero subdeterminants
in the neutrino sector,
is determined by the
left-handed charges \cite{MG}.}.

$\bullet$
The (1--2) mixing relevant to the solar neutrino oscillations,
is dominated by the mixing in the charged lepton sector, and is
of ${\cal O}(\bar{\epsilon}^2)$, in good agreement with the
small angle MSW solution.

$\bullet$
The ratio of the two heaviest 
eigenvalues is ${\cal O}(\bar{\epsilon}^{3})\simeq
10^{-2}.$ Thus if the heaviest neutrino has mass $0.1~{\rm 
eV}$, consistent with
atmospheric neutrino oscillation, the next neutrino will have mass $%
{\cal O}(10^{-3} ~{\rm eV}).$ Given the uncertainties due to 
 ${\cal O}(1)$ coefficients that may not be
predicted by an Abelian symmetry, this 
is certainly in the mass range needed to generate solar neutrino
oscillations via the small-angle MSW solution.

$\bullet$
In our example, we get large hierarchies between $%
m_{eff}^{22}$ and $m_{eff}^{33}$, due to the splittings
in $m^D_\nu$ and $M_{\nu_R}$.
This means that our solutions are stable under 
renormalisation-group effects, even
for large $\tan\beta $.

{\bf (F)~} 
Let us now briefly quote what is the simplest
expectation for  the {\bf Pati-Salam} group
($SU(4)\times SU(2)_L \times SU(2)_R$)
\cite{pati}.
Under $SU(4)$, all the the left(right)-handed
fermions of a given generation, belong in the same 
representation of the group.
The most natural expectation therefore
is that the situation resembles the one in $SO(10)$, 
with the additional freedom
that the mass matrices can be asymmetric. 
However (since
the mass matrices for the quarks and leptons have identical
structure), in the simplest realisation
we need the help of Clebsch factors \cite{CL}
in order to reconcile the observed quark and lepton hierarchies.

\section{An outline of a string-embedded example}

We would now like to look at how the above GUT
analysis may be extended for string-embedded
grand-unified models. In such a case, 
the following generic comments can be made:

$\bullet$
In such a model, the $U(1)$ symmetries are
specified from the string and one generically
expects a product of abelian groups, rather
than a single $U(1)$.

$\bullet$
There are many singlet fields
involved in the mass generation,
not just $\theta$, $\bar{\theta}$. However, their
quantum numbers are specified and the
possible solutions are thus constrained. Moreover,
the field vevs that determine the
magnitude of the various entries, 
are also constrained by the flat
directions of the theory.

$\bullet$ 
Additional string symmetries
(expressed through the string selection
rules) further constrain
the possible forms of the mass matrices, since they
forbid most of the Yukawa couplings that are 
allowed by the rest of the symmetries of
the model.

As an example of the above we consider the 
string-derived flipped $SU(5)$ model \cite{aehn}, 
working with the mass matrices
discussed in~\cite{ELLN,ELLN2}.
Looking at the field assignment in group representations,
one sees that:
(i) since the charge conjugate of the right-handed neutrinos have the same
charge as the down-quarks the Majorana mass matrix will be constrained by
this charge assignment.
(ii) Moreover, due to the above charge
assignments, the Dirac neutrino mass matrix is the transpose of the up-quark
mass matrix.
The quark and charged-lepton mass matrices have been presented
in \cite{ELLN}, where we also reconsidered the possible flat
directions of the theory. 
Since the analysis of the surviving couplings
after all symmetries and string selection rules
are taken into account is quite involved,
we refer for
the details to the original references, while here we just give
an illustration of the predictions for neutrino masses.

The Dirac neutrino mass matrix, $m_{\nu}^D$,  is expressed
in terms of three expansion parameters
\begin{equation} 
m_{\nu}^D = 
\left (
\begin{array}{ccc}
x f & 1 & 0 \\
f & x & 0 \\
0 & 0 & y
\end{array}
\right) 
\label{Dira}
\end{equation}
where the involved fields obey the following
constraints: $x$ is a combination of 
hidden-sector fields that transform as
sextets under $SO(6)$ and needs to be
${\cal O}(1)$ for realistic quark mass
matrices \cite{ELLN}.
$y$ stands for the $SU(5)$ decuplets that break
the gauge group down to
the Standard Model, with vev
$\approx M_{GUT}/M_{s}$,
$M_s$ being the string scale. In weakly-coupled string constructions,
this ratio is suppressed; however in
 the strong-coupling limit of
M-theory, the GUT and the string scales can coincide
and then $y \approx 1$. Finally,
$f$ stands for a singlet field with
vev $\approx 0.04$, again fixed from
the quark hierarchies.

In \cite{ELLN2}, where the expectations
for neutrino masses were studied, we
ended up with two possible forms
for $M_{\nu_R}$, depending
on the vev of the singlet fields. These were
\bea
M_{\nu_R} \propto
\left (
\begin{array}{ccc}
M & 0 & 0 \\
0 & 0 & f y \\
0 & f y & t x
\end{array}
\right )
\label{maj1}
\eea
and 
\begin{equation} 
M_{\nu_R} \propto 
\left (
\begin{array}{ccc}
f y^2  & 2 x y^2  & 0 \\
2 xy^2  & 0 & f  y \\
0 & f y & t x
\end{array}
\right) \label{maj2}
\end{equation}
where in the second example the factor of $2$ has been included
so as to avoid {\em sub-determinant cancellations},
which are not expected to arise once 
coefficients of order unity are properly taken into account.

As we see, in the first matrix the lightest right-handed neutrino
decouples from the rest, while in the second there is
large mixing involving all neutrinos. In the next
section, we will see how the two structures lead to
completely different predictions for leptogenesis.
Here, we would simply like to stress that
the  potentially large off-diagonal entries 
in the heavy Majorana mass matrix 
may yield large neutrino mixing.  Moreover, 
the neutrino Dirac matrix 
also provides a potential source of 
large $\nu_{\mu}$ --$\nu_{e}$ mixing. 

In the solution corresponding to the matrix
(\ref{maj1}),  consistency with the neutrino data
implied
$y \approx 1 $ (as could occur in the strong-coupling limit 
of M-theory)
and  $ t \sim f$.
The actual value of
$M$ was found to be irrelevant, provided
$M$ is not anomalously small, thus increasing $m_{1}$ to
an unacceptable value.

For the second example, 
 $m_{eff}$  is given by
\bea
m_{eff} \propto \left ( 
\begin{array}{ccc}
-f^4 x^2 + tfx & -f^4 x - tf x^2 & -f^2 y^2  \\
 -f^4 x - tf x^2 &  -f^4 - 3 f t x^3 & f^2 x y^2 \\
-f^2 y^2 & f^2 x y^2 & -4 x^2 y^4 
\end{array}
\right )
\eea
and now 
the solutions with large light neutrino
hierarchies
require  $y \approx f$
and  $ t \leq f^3$, so that
the entries in the (1,2) sector of $m_{eff}$
remain small. 

We therefore see that the observed neutrino data 
are reproduced in this model. This is important,
given the constrained form of the various operators,
due to  the string selection
rules. Moreover, the requirement for matching
the observations constrains
the model parameters. In the next section we
are going to see what additional information
we may obtain on the structure of the theory
from leptogenesis. Although we will use this
model as a guideline, the results are more
generic and apply to a wide range of neutrino
textures.

\section{ Leptogenesis and Neutrino Textures}

In our previous analysis we saw that in many models
the neutrino masses  are largely of the Majorana type, 
implying the existence of
interactions that violate lepton number.
Then, it is natural to wonder whether such masses can
have any cosmological implications.
Among various proposals, it has been pointed out that 
the out-of-equilibrium
decay of heavy Majorana neutrinos
may lead to a net lepton asymmetry in the
universe \cite{fyan}, which is converted to a
baryon asymmetry at the electroweak phase
transition \cite{MIS}.

What are the constraints on Super-Kamiokande friendly textures,
that we can derive from leptogenesis? 
In what follows, we will only outline the 
basic points of the discussion that appeared
in \cite{ELLN3}.
Let us  start by considering the case where there is a hierarchy
of eigenvalues in the 
heavy Majorana neutrino mass matrix,
$M_{N_1} < M_{N_2},M_{N_3}$.
In this case, 
any lepton asymmetry is generated by the $CP$-violating decay of
    the lightest right-handed neutrino $N_1$. 
At tree level, the total decay width of $N_1$ 
(for  both the modes
$N_1 \rightarrow \phi^\dagger + \nu$
and $N_1 \rightarrow \phi + \bar{\nu}$,
where $\phi$ is the Higgs field)
is given by
    \beq
\Gamma = \frac{(\lambda^\dagger \lambda)_{11}}{8 \pi} M_{N_1}
    \eeq
where $\lambda = m^D_\nu /v$, $v$ being the corresponding
light Higgs vev. As usual,
the leading contribution to
the $CP$-violating decay asymmetry, arises from the
interference between the tree-level decay amplitude and one loop amplitudes.
These include corrections of vertex type, 
but may also involve self-energy corrections \cite{rev}.

What is the allowed structure of the heavy Majorana matrices?
In a cosmological model with inflation, the decays of the right-handed neutrinos
should occur below the scale of inflation (which
is constrained by the magnitude of the density fluctuations
observed by COBE). This gives
\bea
M_{N_i} \leq m_{\eta} \leq 10^{13} ~\hbox{GeV}
\label{inflaton}
\eea
where $m_\eta$ is the inflaton mass (note that this
bound may be increased in models with preheating).
Then, incorporating the constraints on $M_{\nu_R}$
from the neutrino data, one finds that
solutions with degenerate neutrinos
require
$M_{N_3}  \approx {\cal O}({\rm a ~few ~times ~10^{13}} ~ {\rm GeV})$.
On the other hand, solutions 
with large light neutrino hierarchies
require 
$M_{N_3} \approx O({\rm a ~few ~times ~10^{14}-10^{15} ~{\rm GeV}})$.
In the latter case, the inflaton mass condition demands
heavy Majorana hierarchies of the type
\begin{eqnarray}
\frac{M_{N_1}}{M_{N_3}}  \leq {\cal O} 
\left (
\frac{1}{100} \right ) <  \frac{M_{N_2}}{M_{N_3}}
\end{eqnarray}
In both cases, the inflation constraint may be
accommodated with ease in models with neutrino
masses that match the experimental data.

The strongest bounds however, arise from the 
requirement that at the time of their decays, the neutrinos
have to be out of equilibrium. 
This implies that the decay rate $\Gamma$ has 
to be smaller than the Hubble parameter $H$ at temperatures $T\approx
M_{N_1}$.  $H$ is given by $
H \approx 1.7 ~g_*^{1/2} ~\frac{T^2}{M_p}$,
where  in the Minimal Supersymmetric Standard model
$g_* \approx 228.75$ (while $g_* = 106.75$ for the Standard
Model), indicating that
\bea
\frac{(\lambda^\dagger \lambda)_{11}}{14 \pi g_*^{1/2}} M_{p} < M_{N_1}
\label{outofeq}
\eea
However, a more accurate constraint is obtained by
looking directly at the solutions of the Boltzmann
equations for the system, since it turns out 
that even for Yukawa couplings larger than indicated in
(\ref{outofeq}),
the lepton-number-violating scatterings mediated
by right-handed neutrinos do not wash out completely 
the generated lepton asymmetry 
at low temperatures \cite{KolbTur,evol}. 
For example, it turns
out that for $M_{N_1} =10^{13}$ and 
$(\lambda^\dagger \lambda)_{11} = 1.6 \times 10^{-3}$,
a suppression factor of $\approx 0.1$ is obtained
\cite{ELLN3}, with respect to the case  where
$\Gamma \ll H$.

To understand the nature of this bound, we also 
have to look at the neutrino Dirac mass matrix. 
For the calculation of the lepton asymmetry, 
we need to know the combination $(\lambda^{\dagger} \lambda)_{11}$.
In the case that the light entry of
$M_{\nu_R}$  decouples from the rest, 
this can be read directly from 
$(m_{\nu}^{D\dagger}  m_{\nu}^D)_{11}$, where
\bea
(m_{\nu}^{D\dagger}  m_{\nu}^D) \propto 
\left (
\begin{array}{ccc}
f^2 (1+x^2) & 2 f x & 0 \\
2 f x & 1+x^2 & 0 \\
0 & 0 & y^2
\end{array}
\right)
\label{DiraSq}
\eea
Since $f \approx 0.04$, 
$(\lambda^{\dagger}  \lambda)_{11}$ is suppressed
and can be compatible with the out-of-equilibrium
condition.

However, let us now consider the second form
of $M_{\nu_R}$, where the lightest eigenvalue
of $M_{\nu_R}$ does
not decouple. Then, in order to work with
the $ M_{N_i}$ mass eigenstates, we need to 
diagonalise $M_{\nu_R}$ and also
transform $m_{\nu}^D$  to the  basis where
$M_{\nu_R}$ is diagonal.
Indeed, let
\bea
M_{\nu_R}^{diag} = V^T \cdot M_{\nu_R} \cdot V
\eea
Then the Yukawa couplings have to be calculated from the
matrix
\bea
\tilde{m}_{\nu}^D ~=~ {m}_{\nu}^D \cdot V
\eea

The mixing matrix, for the
field vevs that were compatible with the
neutrino data and for the 
coefficient choice of eq.(\ref{maj2}) is
given by
\bea
V \approx 
\left (
\begin{array}{ccc}
0.63 & 0.63 & -0.45 \\
0.70 & -0.70 & 0.18 ~f \\
0.32 & 0.32 & 0.89 
\end{array}
\right )
\eea
This leads to
\bea
\tilde{m}_{\nu}^D ~\approx~ 
\left (
\begin{array}{rrr}
0.70 & - 0.70 & -0.27 f \\
0.70 & -0.70 & -0.27 f \\
0.32 f & 0.32 f & 0.9 f 
\end{array}
\right )
\eea
indicating that
$(\lambda^{\dagger} \lambda)_{11}$ 
may never  be consistent with the out-of-equilibrium conditions.

The above leads to the following generic conclusion:
Models with 
(i) a small 1--2 mixing in the heavy Majorana
mass sector or (ii) small (1,1) and (1,2) couplings in the 
Dirac-neutrino 
sector, are naturally compatible with leptogenesis.
On the other hand, if a non-trivial 1--2 mixing
in $M_{\nu_R}$ gets combined with large off-diagonal
entries in the Dirac-neutrino sector,
the out-of-equilibrium condition for leptogenesis
tends to get violated.

\section{ Lepton-number violation in
Ultra-High-Energy neutrinos}

Besides neutrino oscillations,
alternative possibilities for explaining the atmospheric
neutrino data have been discussed, such as neutrino
decay \cite{Pak} and flavour--changing neutrino--matter interactions
\cite{GonzaG}. The latter 
(which arise in many Standard Model extensions, such as
$R$-violating supersymmetry and leptoquark models)
had already been used 
in the past, for solar neutrino conversions.
However, in recent proposals it has been shown that
they may also account for the super-Kamiokande
observations, without directly discussing neutrino masses.
The relevant process would be
$\nu_\mu + f \rightarrow \nu_{\tau} + f$
where the required couplings are of the order of
$\lambda_{\tau f} \cdot \lambda_{\mu f} \approx 0.1$,
for propagators with masses of 200 GeV \cite{GonzaG}.
Then, the immediate question that arises is whether
there is any way to directly probe such couplings.
In this framework, it had already been pointed out
that  such couplings  may induce significant changes
in the interaction rates of ultra-high energy neutrinos (UHE)
with nucleons and electrons,
through the production of particle resonances \cite{UHE}.

To make the analysis more specific, we 
will discuss lepton-number violation in
the framework of $R$-violating SUSY, however, the
results are more generic.
In these models the lepton-number violating operators
that are consistent with the symmetries of the theory are
\bea
W_{\Delta L \neq 0}  =  \lambda_{ijk} L^i L^j \bar{E}^k 
+\lambda^{\prime}_{ijk} L^i Q^j \bar{D}^k 
\eea
where $i,j,k$  are generation indices. $L^i \equiv 
(\nu^{i},e^{\,i})_{\mathrm{L}}$ and  
$Q^i \equiv (u^{i},d^{\,i})_{\mathrm{L}}$ are the left-chiral
superfields, and 
$\bar{E}^i \equiv e^{\,i}_{\mathrm{R}}$, 
$\bar{D}^i \equiv d^{\,i}_{\mathrm{R}}$, and 
$\bar{U}^i \equiv u^{i}_{\mathrm{R}}$ 
the right-chiral ones.
The good agreement between the data and 
the standard-model expectations implies 
bounds on the strength of lepton-number-violating
operators \cite{rviolb}.
Then, one has the following:
$LQ\bar{D}$-type interactions 
of electron neutrinos or antineutrinos with the first-generation quarks 
are highly constrained from
various processes, such as
neutrinoless double-beta decay, 
charged-current universality, atomic parity violation,
and the decay rate of $K \rightarrow \pi \nu \bar{\nu}$.
The bounds on $LL\bar{E}$
 couplings are also relatively strong.
On the other hand,
experimental limits on the 
$L^iQ^j\bar{D}^k$ couplings that involve $\nu_{\mu}$,
which would be relevant to explaining the
Super-Kamiokande data, are  less restrictive.
Some useful bounds in the case that one
$R$-violating coupling dominates appear in Table 3.

\vspace*{0.2 cm} 

\begin{center}
\begin{tabular}{|c|c|}
\hline \hline
Coupling & Limited by  \\
\hline \hline 
$\lambda_{12k} < 0.1~(2\sigma)$ & charged-current universality  \\
  $\lambda_{131, 132, 231} < 0.12~(1\sigma)$ &  $\Gamma(\tau\rightarrow 
  e\nu\bar{\nu})/ \Gamma(\tau \rightarrow \mu\nu\bar{\nu})$  \\
  $\lambda_{133}< 0.006~(1\sigma)$ & $\nu_{e}$ Majorana mass \\[6pt]
  $\lambda^{\prime}_{21k} < 0.18~(1\sigma)$ & $\pi$ decay  \\
  $\lambda^{\prime}_{221} < 0.36~(1\sigma)$ & $D$ decay  \\
  $\lambda^{\prime}_{231} < 0.44~(2\sigma)$ & $\nu_{\mu}$ deep 
  inelastic scattering  \\
       \hline \hline
\end{tabular}
\end{center}  
{ \small Table 3: {\it Experimental constraints 
(at one or two standard deviations) on 
the $R$-violating Yukawa couplings of interest, for the case of 
200-GeV sfermions.  For arbitrary sfermion mass 
the limits scale as by $(m_{\tilde{f}}/200~{\rm GeV})$, except for 
$\lambda^{\prime}_{221}$.}}

\vspace*{0.3 cm} 

UHE neutrinos are produced from the 
interactions of energetic protons in active galactic 
nuclei (AGN), as well as 
from  gamma-ray bursters or
pion photoproduction on the cosmic microwave background.
Moreover, they
may also arise from exotic heavy-particle
decays and the collapse of topological defects.
Their effects can be observed in neutrino 
telescopes \cite{chpr} and, in this respect it is important to
look for specific signals of lepton-number violation
as km$^{3}$-class neutrino observatories come into being.
The dominant mechanisms for producing UHE
photons and neutrinos are expected to be
\dk{p\:(p/\gamma)}{\pi^{0}+ {\rm anything}}{\gamma\gamma}
and
\dkdk{p\:(p/\gamma)}{\pi^{\pm}+ {\rm
anything}}{\mu\nu_{\mu}}{e\nu_{e}\nu_{\mu}\;.}
If $\pi^{+}$, $\pi^{-}$, and $\pi^{0}$ are produced in equal numbers, 
the relative populations of neutral particles will be 
$2\gamma:2\nu_{\mu}:2\bar{\nu}_{\mu}:1\nu_{e}:1\bar{\nu}_{e}$. 
Since there  are no significant conventional sources of $\nu_{\tau}$ 
and $\bar{\nu}_{\tau}$, we are not able to
probe lepton-number-violating operators of the
$L_3 Q \bar{D}$ type.

What is the effect of the new couplings? 
Let us first consider 
$\nu_{\mu}N$ interactions.
The charged-current reaction 
$\nu_{\mu}N \rightarrow \mu^{-}+{\rm anything}$ can receive 
contributions from (i) the $s$-channel process           
$\nu_{\mu}d_{\mathrm{L}}\rightarrow \tilde{d}^{k}_{\mathrm{R}} 
\rightarrow \mu^{-}_{\mathrm{L}}u _{\mathrm{L}}$, which involves 
valence quarks, and from 
(ii) $u$-channel exchange of 
$\tilde{d}^{k}_{\mathrm{R}}$ in the reaction $\nu_{\mu}\bar{u} 
\rightarrow \bar{d}\mu^{-}$, which involves only sea quarks.  
As a  consequence of the spread in quark momenta, the resonance peaks 
in case (i) are not narrow, but are
broadened and shifted above the threshold energies.
The right-handed squark $\tilde{d}^{k}_{\mathrm{R}}$ has a similar 
influence on the neutral-current reaction $\nu_{\mu}N \rightarrow 
\nu_{\mu}+{\rm anything}$.  On the other hand, 
left-handed squarks can contribute {\em only} 
to the neutral-current reaction and we therefore 
predict modifications to
the ratio of neutral-current to charged-current interactions
\cite{UHE}.
Similar effects are observed in 
$\bar{\nu}_{\mu}N$ interactions.
In Figure \ref{RATIO},we compare the ratio 
$\sigma_{\mathrm{NC}}/\sigma_{\mathrm{CC}}$ in the standard 
model with the case where lepton-number-violating
couplings are present.
In this calculation, we use the CTEQ3 parton distributions \cite{CTEQ3}.
Although neutrino 
telescopes will not distinguish 
between events induced by neutrinos and antineutrinos and 
the relevant quantity would thus be the sum 
of the $\nu_\mu N$ and $\bar{\nu}_\mu N$ cross 
sections, we present these processes separately in order to
stress the effects of the helicity structure of the theory.

\begin{figure}
\centerline{
\psfig{figure=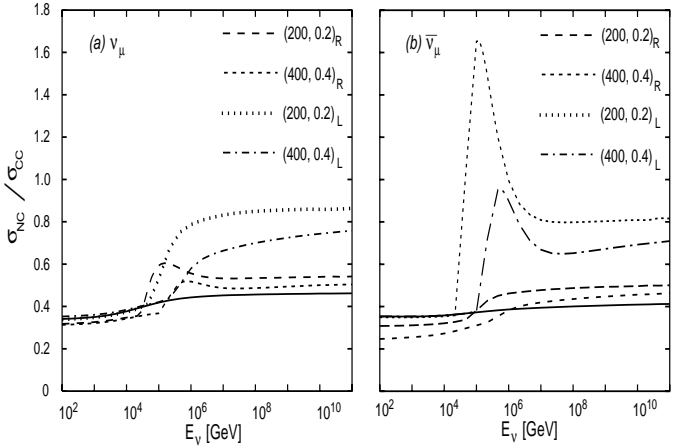,width=9.0cm,height=6.0cm,angle=0}
}
\caption{\it 
Neutral-current to charged-current ratios for 
        (a) $\nu_{\mu}N$ and (b) $\bar{\nu}_{\mu}N$ 
        interactions.  The solid lines show the predictions of the standard 
        model.  The dashed (short-dashed) curves include the contributions of 
        a right-handed squark, $\tilde{d}^{k}_{\mathrm{R}}$, with mass 
        $\widetilde{m}=200~(400)~{\rm GeV}$ and coupling 
        $\lambda^{\prime}_{21k}= 0.2~(0.4)$.  The dotted (dot-dashed) curves 
        include the contributions of a left-handed squark, 
        $\tilde{d}^{k}_{\mathrm{L}}$, 
        for the same  masses and couplings.}
\label{RATIO}
\end{figure}

We see that the modifications from the 
standard-model cross sections are appreciable,
even away from the resonance bump. 

What about neutrino interactions on electron targets?
In the standard model,
because of the smallness of the electron-mass, neutrino-electron interactions 
in matter are weaker than neutrino-nucleon 
interactions, with the exception of the
 resonant formation of the intermediate boson in 
$\bar{\nu}_{e}e \rightarrow W^{-}$ interactions \cite{chpr}.
Additional effects may arise through $R$-violating interactions
\cite{UHE}.
Because the $LL\bar{E}$ couplings are constrained to be 
small, only channels that involve resonant slepton production can 
display sizeable effects. 
Such couplings are too small to explain the Super-Kamiokande
data in the framework of \cite{GonzaG}, nevertheless it is interesting
to investigate whether they could have any observable effect.
Small couplings result in small decay widths,
and consequently, 
it will be difficult to separate such 
a narrow structure from the standard-model background.
One  interesting characteristic is that the slepton resonance will only 
be produced in downward-going interactions. Indeed, in 
water-equivalent units, the interaction length is given by
\begin{equation}
        L_{\mathrm{int}}^{(e)} = \frac{1}{\sigma(E_{\nu})(10/18)
        N_{\mathrm{A}}}\; 
        \label{eq:lint}
\end{equation}
where $N_{\mathrm{A}}$
is the Avogadro's number 
and $(10/18)N_{\mathrm{A}}$ is the number of electrons in a mole of 
water.  At the peak of a 200 (400)
GeV slepton resonance 
produced in $\bar{\nu}e$ interactions, the 
interaction length indicates that the resonance is effectively 
extinguished for neutrinos that traverse the Earth.

Still, it would be easier to observe a slepton resonance in the
case where the produced final states clearly stand out above the 
background. One such possibility arises if many $R$-violating couplings
are simultaneously large, thus leading to exotic final-state topologies.
An even better possibility arises if 
neutralinos are relatively light. In this case, the 
slepton may also decay into the corresponding 
lepton and a light neutralino, which  in its turn decays into leptons 
and neutrinos:
\dkp{\nu_{\mu}e^{-}_{\mathrm{L}} \rightarrow 
\tilde{\tau}^{-}_{\mathrm{R}} 
}{\tau^{-}}{\tilde{\chi}^{0}}
{\tau^{+}_{\mathrm{R}}\nu_{e}\mu^{-}_{\mathrm{L}}\hbox{ or }
\tau^{+}_{\mathrm{R}}\nu_{\mu}e^{-}_{\mathrm{L}}}
and
\dkp{\bar{\nu_{e,\mu} }e^{-}_{\mathrm{R}} \rightarrow 
\tilde{\tau}^{-}_{\mathrm{L}}\rightarrow}{\tau^{-}}{\tilde{\chi}^{0}}
{\tau^{+}_{\mathrm{L}}\bar{\nu} e^{-}_{\mathrm{R}}\;}

The decay length of a 1-PeV $\tau$ is about 50 m, 
so the production and subsequent decay of a $\tau$ at
UHE will result in a characteristic ``double-bang'' signature 
in a Cherenkov detector.  
Because there are no conventional
astrophysical sources of tau-neutrinos, 
while 
$\tau$-production through a 
slepton resonance with a mass $\geq$ 200 GeV, is essentially
background-free,
reactions 
that produce final-state $\tau$-leptons are 
of special interest for probing new physics.

\section{Summary and Conclusions}

We discussed aspects of neutrino masses and lepton-number violation, in the
light of the observations by Super-Kamiokande. 
We first studied phenomenological textures
which match the data from various experiments
and then investigated how such structures may arise,
in models with flavour and GUT symmetries.
In supersymmetric extensions of the Standard Model, renormalisation 
group effects were found to be important. In particular, 
for small $\tan\beta$, $b$--$\tau$ unification requires the presence of 
significant $\mu$--$\tau$ flavour mixing.  
On the other hand, for large $\tan\beta$, 
very small mixing at the GUT scale may be amplified to 
maximal mixing at low energies, and vice versa. 
Leptogenesis may give additional constraints on
neutrino mass textures. Channels to directly search 
for lepton-number violation in ultra-high energy neutrino 
interactions, have also been proposed.

\acknowledgements{}
I would like to thank John Ellis, G.K. Leontaris, D.V. Nanopoulos
and G.G. Ross for very stimulating collaborations
that led to the work that has been presented on
the implications of the Super-Kamiokande  data.
I also thank M. Carena, D. Choudhury, H. Dreiner,
G.K. Leontaris, C. Quigg, G.G. Ross, C. Scheich and
J.D. Vergados, for earlier, equally
stimulating collaborations on various aspects of neutrino physics.
Financial support from the Corfu Summer Institute
is gratefully acknowledged.

\end{narrowtext}
\end{document}